\documentstyle[aps,preprint,tabularx,graphicx]{revtex}
\begin{document}
%
%
\title{Unzipping double-stranded DNA with a force: scaling theory}
\author{Jeff Z.Y. Chen}
\address{Department of Physics,
University of Waterloo, Waterloo,
Ontario,Canada}
\date{\today}
\maketitle
\begin{abstract}

A double stranded DNA molecule under the stress of a pulling force
acting on the strand terminals exhibits a partially denatured
structure or can be completely unzipped depending the magnitude of
the pulling force.
A scaling argument for relationships amongst basic length scales
is presented that takes into account the heterogeneity of the
sequence. The result agrees with our numerical simulation data,
which provides a critical test of the power laws in the unzipping
transition region.

\end{abstract}


\pagebreak

The characteristics of the structure of double stranded DNA
(dsDNA) have been recently probed through mechanical manipulation
at a microscopic level by applying an external
force.\cite{Bock,Marko} From a theoretical modelling perspective,
we have just begun to understand these experimental results, which
are usually provided in the form of manipulating force verses
extension. Most of the current theoretical work is based a
homopolymer model, in which the binding forces between nucleotides
in various types are treated equally \cite{Marko,Bhat,Zhou,Mare}.
One of the theoretical efforts is to reproduce the force-extension
phase diagram observed experimentally, together with other
interesting conformational properties that characterize the
unzipping transition. It is certainly a challenging task to
understand the system-dependent and the universal features in
these complex systems, as the pairing along the two DNA strands is
in general heterogeneous rather than homogeneous\cite {LN}.

The present article serves two purposes. Firstly, a physically
direct scaling argument is employed to analyze a simplified {\it
heterogeneous} dsDNA model, which exhibits an unzipping transition
 under the external pulling force acting upon the terminal end.
  The usefulness of some well-tested scaling arguments
 in polymer theory is demonstrated, both in terms of a Gaussian
 model and of a freely jointed bond model that correctly addresses the finite extensibility of each strand.
Secondly, based on a continuous description of the problem
\cite{LN}, numerical simulations are presented to test some of the
obtained power laws, in particular those near the unzipping
transition. A special attention is paid to display the
sequence-dependent features verses averaged, sequence-independent
features.

To start with, we consider two bounded DNA strands as shown
schematically in Fig. 1, where some segments interact with each
other predominantly with repulsions by displaying excursion loops,
and some segments predominantly with attractions in a mostly bound
form. To simplify the consideration, we assume here that the
interaction energy per base pair when they are confined in the
bounding distance can either be $+\epsilon$ (repulsive) or
$-\epsilon$ (attractive). Furthermore, we assume that the averaged
free energy per base pair is $-|f_0|$ when the dsDNA is below the
melting temperature with no additional pulling force.

When an external pulling force is applied to the two terminal
nucleotides (or two terminal groups) of the dsDNA, we assume that
on average $M$ base units can be pulled out. Hence, the reduced
free energy of the entire chain reads
\begin{equation}\label{Ftot}
 \beta {\cal F}  = -(N-M)\beta|f_0| + {D\over 2} {{Z^2}\over{Ma^2}} -
\beta FZ -  {\sqrt M} \beta\epsilon .
\end{equation}
where the first term describes the fact that there are $N-M$ base
pairs that still remain in the bound state on average. The three
additional terms describe the effects of the pulling process that
results in $M$ separated base pairs and an average distance $Z$
between the two terminals. One of the essential assumptions here
is that the polymer segment in the stretched portion obeys a
Gaussian distribution\cite {DG}, an assumption that will be
questioned below. In order to make a more precise comparison with
a previous result, a prefactor $D/2$ is explicitly introduced in
the second term, to make it exact even at the level of numerical
coefficients. The third term is simply a potential energy
reduction of the terminal pairs when an external pulling field is
applied.

The last term in Eq. \ref{Ftot} describes the binding energy
reduction of the pulled portion. When the dsDNA is being pulled,
two possibilities might occur near the end of the separated
portion. In the event that this portion is connected to an
originally bound segment, a stronger force is required to unzip
this locked portion. In the event that this portion is connected
to a segment that is an excursion loop, the dsDNA would
spontaneously unwind the excursion loop. Hence under a fixed
pulling force and distance $Z$, $M$ may vary and the entire
separated portion of the $M$ base pairs would adjust itself so
that it has a net sum of a lower energy. Since the binding energy
is fluctuating between $+{\sqrt M}\epsilon$ and $-{\sqrt
M}\epsilon$ {\it randomly} the average energy that the pulled
portion can acquire is $-{\sqrt M}\epsilon$, where the square root
is a result of the randomness in the system\cite {IM}.

Now the free energy has to be minimized with respect to $Z$ and
$M$ as adjustable parameters. Minimizing with respect to $Z$ leads
to
\begin{equation}\label{zeq}
Z = a^2 \beta FM /D
\end{equation}
which has a linear form in $F$ and reflects the fact that the
pulled portion is near a highly stretched limit. Inserting Eq.
\ref{zeq} into Eq. \ref{Ftot}, we have
\begin{equation}\label{FM}
{\cal F} = - N|f_0| + tM - {\sqrt M} \epsilon
\end{equation}
where we have defined a reduced force parameter
\begin{equation}\label{tdef}
t = |f_0| - \beta F^2a^2/(2D).
\end{equation}

The above equation is our final expression for the free energy of
the system. We see that when $t\le 0$, there is no finite minimum
in the free energy as $M$ tends to increase unboundedly; as a
consequence the dsDNA is entirely unzipped. When $t > 0$ however,
there is an equilibrium state, where, by minimizing the above
expression with respect to $M$, the average number of unzipped
base pairs is
\begin{equation}\label{Meq}
 M = (\epsilon /t)^2/4
\end{equation}
and the free energy is
\begin{equation}\label{ft}
{\cal F} = - N|f_0|- \epsilon^2/t/4.
\end{equation}
The unzipping transition takes place when $t=0$, which is
equivalent to saying that there exists a critical unzipping force,
\begin{equation}\label{fc}
F_{\rm c} =  {\sqrt {2D|f_0|/(\beta a^2)}}. $$
\end{equation}

The main conclusion drawn here, namely, Eqs. 5, 6, and 7, is
consistent with a much more involved derivation\cite{LN}. The
numerical coefficient of the critical force is exact, in view of
the fact that the original involved terms in the free energy are
exact to start with. 
The free energy for the entire system is no longer self-averaging,
as the last term in Eq. \ref{ft} is not proportional to $N$.
Indeed, from the physical perspective, the equilibrium state of
the pulled portion is established regardless of the total length
of the dsDNA, as long as the pulled portion is still shorter. The
independence of $N$ in the last term is merely a statement that
for different $N$, the equilibrium state of the pulled portion
remains essentially the same.

This simple treatment for the unzipping transition of the dsDNA
produces several relationships between the basic length scales and
the transition properties. However, there is one important feature
of the stretched portion that has been ignored both above and in
Ref. \cite{LN}, that might become a pitfall of the entire
derivation. The usage of a Gaussian entropic term for the
stretched portion in Eq. 1 can be traced back to the spring-bead
model commonly used in polymer modelling. As a spring can be
stretched without a bound under a strong pulling force, the
current model physically allows the pulled segment to stretch
beyond the natural length $Ma$, as demonstrated in Eq. (2). Hence,
it becomes desirable to incorporate the effect of finite
extensibility in the above discussion.

It has been previously demonstrated by Grosberg and Khokhlov
\cite{GK} that a better representation for a stretched polymer is
the freely jointed bond model that contains a rigid bond that
models a polymer monomer unit. Instead of the second and thrid
terms in Eq. 1, the contribution to the free energy due to the
entropic penalty in the strong stretched limit ($D=3$) can be
shown to be
\begin{equation}\label{GK}
{\cal F}_{\rm stretch} = {M\over\beta}{\rm ln} [{{{\rm sinh}
(\beta Fa)}\over {\beta Fa}}]
\end{equation}
and the separation distance can be directly calculated,
$$
Z = Ma  [{\rm coth} (\beta Fa) - 1/(\beta Fa)]. \eqno (9)
$$
Replacing the the second and the third terms in Eq. 1, we have
$$
{\cal F}  = -(N-M)f_0 + {M\over \beta} {\rm ln} [{{{\rm sinh}
(\beta Fa)}\over {\beta Fa}}] - {\sqrt M} \epsilon\eqno(10)$$ The
linear dependence of the stretching free energy on $M$ essentially
renders the same scaling theory as discussed above.
Indeed, 
the free energy expression in Eq. 6 remains the same and the only
change is the definition of $t$, which now reads
$$
t = |f_0| + {1\over \beta}{\rm ln} [{{{\rm sinh} (\beta Fa)}\over
{\beta Fa}}] .\eqno (11)
$$
We note that this expression recovers the same form as in Eq. 4 in
the small $\beta Fa$ limit.

Hence, when the finite extensibility is taken into account in the
dsDNA model, the critical force can be implicitly solved through
finding the root for $t = 0$ in Eq. 11. The power laws in Eqs. 5
and 6 remain the same.

The rest of this article is devoted to the discussion of numerical
simulations that can be used to address system specific features
in comparison with these universal features. Returning to the
Gaussian description of the conformation for both strands of DNA,
we write a more rigorous expression for the probability functional
in one dimension, after Lubensky and Nelson (LN)\cite {LN}:
$$
P_0  \propto {\rm exp} \{ - {1\over {2 a^2}} \int_0^N [
({{dz}\over {ds}})^2 ] ds - \beta \int_0^N W[z(s), s] ds \} \eqno
(12)
$$
where $z(s)$ describes the distance between the base pair labeled
$s$ in the dsDNA, and $W$ is a strong bounding potential, which
depends on the (disordered) sequence of nucleotides on both
strands. In a Gaussian chain model, the Kuhn length $a$ in this
reduced probability function is $\sqrt{2}$ times the Kuhn length
of each single strand of DNA \cite{note2}. The unzipping
probability is then simply
$$
P_{\rm unzip} [z(s), \{s\}] = P_0[z(s), \{s\}] {\rm exp} \{ \beta
{ F} \cdot z(N) \} \eqno (13)
$$
with the assumption that the pulling force is directed along the
$z$ direction and acting on the $N$th monomer. The $\{ s \}$
dependence in these functions stresses the fact that the
interaction between the base pairs depends on a given sequence.

A mathematically similar problem to the above probability
distribution function arises in the study of random copolymer
localization in an external field, specifically, at a sharp
interface\cite {Garel}.The basic idea of our numerical simulations
is to simulate the disordered heterogeneousness in the system by
generating an ensemble of dsDNA models that have explicit bounding
interaction dependence. The calculation of the probability
function of the LN model is then translated into solving a Schr{\"
o}dinger type equation. For each generated sequence, the Schr{\"
o}dinger equation is solved and thus yields an exact result with
no arbitrary approximations\cite {Chen}.

We have used basically the same numerical procedure described in
Ref. \cite {Chen} by computing the probability function, $\psi(z,
{s})$, of finding the $s$th base pair having a separation distance
$z$. Now, the calculation of $\psi(z,{s})$ in a potential field
can be simplified to solving a time-dependent Schr{\" o}dinger
equation,
$$
-{{\partial \psi (z, s)}\over {\partial s}} = [ - {a^2\over 2
}{d^2\over {dz^2}} + \beta W(z,  s ) ]\psi (z,s) \eqno (14).$$
where the ``wave function'' $\psi$ is subject to the initial
condition $\psi(z,0) = 1$. The probability function for the
terminal pair to have a separation distance $z$, $q(z, \{ s\})$,
is given by $q(z, \{ s \}) = \psi (z,N)$. At this stage, $q(z, \{s
\})$ has an explicit sequence dependence.

For further numerical development we must choose a function form
for the interaction potential and we write
$$
W[z, \{ s \}] = \zeta(s) \epsilon w(z) \eqno (15)
$$
with $w(z)$ being modelled by a Gaussian potential well $w(z) = -
{\rm exp}(-(z/a)^2) /\sqrt \pi$. The coefficient $\zeta(s)$
represents the binding energy of different pairs with $\epsilon$
assumed here positive. In the remainder of this article we
describe numerical results of having chosen $\beta\epsilon = 1$.

There are in total 10 possible pairs of nucleotides in DNA,
corresponding to 10 different values of $\zeta(s)$, where
guanine-thymine and adenine-cytosine pairings are known to be
attractive. Whether or not the disordered sequence of bases along
DNA is truly {\it random} is a still debatable question. However,
for the present purpose, it suffices to assumed that $\zeta(s)$
takes a statistically independent random value for each $s$.
In particular, we assume 
 $\zeta(s)$ is a random number uniformly
ranging from $-1$ to $1$ as a theoretical abstraction. The
numerical data for a biased randomness will be discussed
elsewhere\cite {unp}.

Figure 2 shows the relationship between the force and the
separated distance $\bar z$ for 10 different, randomly generated
disorder sequences. In producing the simulation data, a finite
system with a limited size of $z<1600$ was considered; this
finite-size effect shows up in all curves as they cannot exceed a
separation distance of 1600. A length of $N=32000$ units has been
used here, which guarantees a minimum finite-size effect due to
the modelled DNA length\cite {note3}. The step-like functions in
this plot are a signature of the sequence disorderliness. This
figure is a clear indication that for a specific dsDNA model (and
indeed for real dsDNA) a variety of force-distance curves can be
obtained with the unzipping transition occur near an averaged
critical force. The critical force for each sequence, however,
varies. For reference, the anticipated critical force, averaged
over the sequence ensemble, is around $\beta F_c a = 0.3724$ from
an accurate estimate of the bound free energy $f_0$ of this system
(see Eq. 7)\cite {unp}.

Though each sequence might show strong individual characteristics,
the universal behavior around which the individual properties
fluctuate, can be obtained by taking an average over the sequence
ensemble. In order to do so, we have produced $10^3$ sample DNA
models, each being examined through the solution of the Schr{\"
o}dinger equation. The average separation, $<\bar z >$, as a
function of the pulling force $F$, is presented in Fig. 2 as the
thick dark curve, where the average $<...>$ represent sampling in
the disorder ensemble.

The same curve is also presented in Fig. 3 by circles as a
function of $F_{\rm c} - F$, in a double logarithmic scale. We see
that within the region of $[0.03, 0.02]$, $<\bar z>$ indeed
follows a power behavior,
$$
<\bar z> \propto (F_{\rm c} - F)^{q_1} , \eqno (16)$$ with a
fitted slope of $q_1=0.194$, directly verifying the scaling
exponent of 2 in Eq. 5. Exception of the scaling can be seen in
smaller $F_{\rm c} - F$ region where the saturation of $<\bar z>$
to a finite value occurs. As a direct consequence of the {\it
compound } finite size effect shown for various sequences in Fig.
2, the averaged $<\bar z>$ is more drastically affected by the
finite size of the system.

LN have quoted two other power laws. Translating the averages  of
$M$ into the averages of $z$ permits us to examine these power
laws:
$$
\delta z_1^2 \equiv <{\bar z}^2 > - <{\bar z}>^2   \propto (F_{\rm
c} - F)^{q_2} \eqno (17)
$$
and
$$
\delta z_2^2 \equiv < {\bar {z^2}}> - <{\bar z}^2 > \propto
(F_{\rm c} - F)^{q_3} \eqno (18)
$$
The square and diamond symbols in Fig. 3 represent  our simulation
data for $\delta z_1^2$ and $\delta z_2^2$. By fitting straight
lines to these curves  in the region $[0.03, 0.02]$, we obtained
$q_2 = 3.99$ and $q_3 = 2.94$, which agree well with analytical
determination of these exponents\cite {LN}.

To summarize, we have carried out a scaling analysis
of the unzipping transition
phenomenon of a heterogeneous dsDNA model. Our main conclusions, on the
free energy and stretching separation,
agree with the results drawn by previous authors. These results compare
favorably well with the numerical simulation data, which, for the first time,
demonstrate the relationship between the sequence-dependent and universal features.

Financial support of this work is provided by the Natural Science
and Engineering Research Council of Canada and the grant associated with the
Premier's Research Excellence Award of Ontario, Canada.



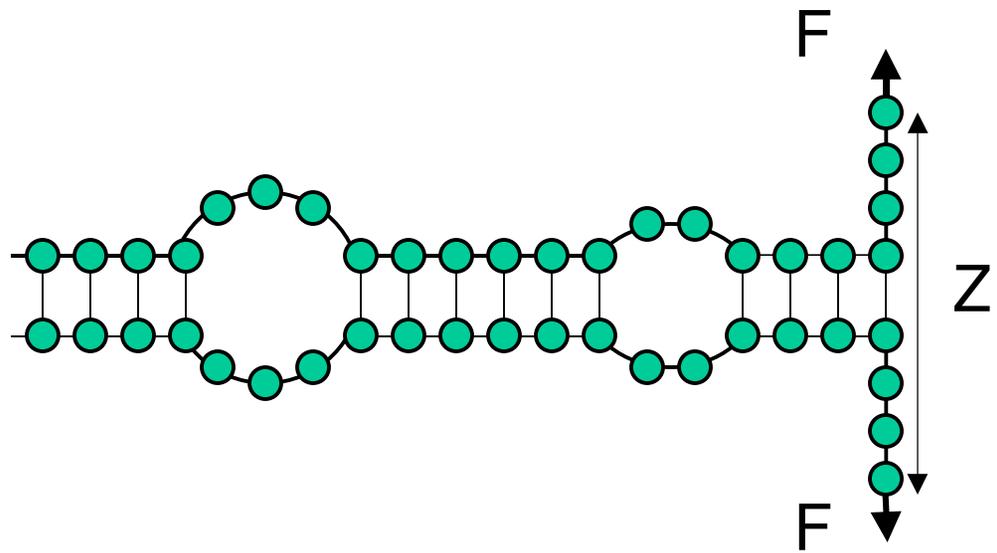
\begin{figure}
\caption{ Schematic representation of the dsDNA model considered.
Each pair interacts  with  either $\epsilon$ or $-\epsilon$, while
the sequence is assumed heterogeneously disordered. }
\label{catoon}
\end{figure}

\begin{figure}
\caption{Typical unzipping force verses extension curves for 10 different
random sequences. The thick curve represents an average over $10^3$ independently generated
samples.}
\label{figure2}
\end{figure}

\begin{figure}
\caption{Scaling behavior of the unzipped extension as a function
of the force difference $F_{\rm c} - F$ (circles). The other two
power laws, Eqs. (17) and (18), are examined in terms of the
numerical data corresponding to squares and diamonds.}
\label{figure3}
\end{figure}








\begin{references}
\bibitem{note2} In reality, dsDNA has a much larger
persistent length in the double helix form, and such features are
not considered here.
\end{references}
\end{document}